\documentclass{aa}  

\usepackage{graphicx}
\usepackage{epstopdf}
\usepackage{dcolumn}
\usepackage{bm}
\usepackage{color}
\usepackage{epsfig}
\usepackage{amsmath}
\usepackage{color}
\usepackage{txfonts}
\def\ee{\end{equation}}
\def\be{\begin{equation}}
\def\bdm{\begin{displaymath}}
\def\edm{\end{displaymath}}
\def\ebe{\end{displaymath}\begin{equation}}

\def\jn2{J_n^2(z)}

\def\l{\left}
\def\r{\right}

\def\bc{\beta _c}
\def\pa{\partial }

\def\bea{\begin{eqnarray}}
\def\eea{\end{eqnarray}}
\def\bd{\begin{displaymath}}
\def\ed{\end{displaymath}}
\def\ba{\begin{array}}
\def\ea{\end{array}}
\def\ebe{\edm\be}

\def\bea{\begin{eqnarray}}
\def\eea{\end{eqnarray}}
\def\ban{\begin{eqnarray*}}
\def\ean{\end{eqnarray*}}
\def\bd{\begin{displaymath}}
\def\ed{\end{displaymath}}

\def\bc{\begin{center}}
\def\ec{\end{center}}

\def\ba{\begin{array}}
\def\ea{\end{array}}

\begin{document}

   \title{Plasma effects on relativistic pair beams from TeV blazars: }

   \subtitle{PIC simulations and analytical predictions}

   \author{I. Rafighi,\inst{1}\fnmsep\thanks{\email{rafighi@uni-potsdam.de}}, S. Vafin\inst{1}, M. Pohl\inst{1,2}, and J. Niemiec\inst{3}
  }

   \institute{Institute for Physics and Astronomy, University of Potsdam, D-14476 Potsdam, Germany
         \and
             DESY, Platanenallee 6, D-15738 Zeuthen, Germany
             \and
Instytut Fizyki J\c{a}drowej PAN, ul. Radzikowskiego 152, 31-342 Krak\'{o}w, Poland
             }

   \date{Accepted August 22, 2017 }

 
  \abstract{Pair beams produced by very high-energy radiation from TeV blazars emit gamma rays in the GeV band by inverse-Compton scattering of soft photons. The observed GeV-band signal is smaller than that expected from the full electromagnetic cascade. This means that the pair beams must be affected by other physical processes reducing their energy flux. One possible loss mechanism involves beam-plasma instabilities that we consider in the present work. For realistic parameters the pair beams can not be simulated by modern computers. Instead, we use a simple analytical model to find a range of the beam parameters that (i) provides a physical picture similar to that of realistic pair beams and (ii) at the same time can be handled by available computational resources. Afterwards, we performed corresponding 2D PIC simulations. We confirm that the beams experience only small changes in the relevant parameter regime, and other processes such as deflection in magnetic field must be at play.   }


   \maketitle
\section{Introduction}\label{Intro}

Many blazars, a sub-class of active galactic nuclei, have been detected with gamma-ray telescopes such as HESS, VERITAS, Fermi and MAGIC as sources of gamma-rays with the energy $E\geq 100$ GeV \citep{Naurois}. These very-high energy photons interact with extragalactic background light (EBL) producing ultra-relativistic electron-positron pairs with the typical Lorentz factor $10^5<\Gamma<10^7$ \citep{RS12a,Miniati13}. The created pairs are subject of many investigations, as they can be affected by several physical processes: (i) inverse Compton scattering (ICS), (ii) deflection by the intergalactic magnetic field (IGMF), or (iii) collective plasma effects. The ICS would result in gamma-ray emission with characteristic energy in the GeV band. But, as indicated by Fermi-LAT data, the GeV gamma-ray emission is suppressed meaning that the ICS is not the fastest of the three processes. The effect of deflection by the IGMF has been well investigated \citep{Neronov09,Neronov10,Taylor11} which led to constraints on the IGMF. However, these constraints are valid only under the assumptions that the multi-TeV gamma-ray emission persists on long timescales and that the pairs lose their energy only due to ICS. The last assumption is very crucial and remains debatable. 

The importance of collective plasma effects has been pointed out by several authors \citep{Broderick12,RS12,Miniati13}. In fact, the pairs can induce electrostatic (two-stream, oblique) and electromagnetic (filamentation, Weibel) instabilities \citep{Breizman74,Breizman90,Bret04,Bret05,Bret06,Bret10,Godfrey75,Lominadze79}. In this case, wave-particle interactions can reduce the energy of the pairs by 30-50 \% \citep{RS02,Bret10}. Therefore, the collective plasma effects can also substantially suppress the GeV-band gamma-ray emission affecting as well the IGMF constraints. 

The pair beams constitute an extremely small fraction of the plasma density in the intergalactic medium (IGM), $\alpha=n_b/n\approx 10^{-16}-10^{-18}$. This circumstance prohibits direct computer simulations of the beams due to insufficient computational power, and substantial adjustments in parameter values have been made in published simulation studies \citep{Sironi14,Kempf16}. 
At the same time, an accurate analytical description of the non-linear evolution of the plasma system is also problematic. In this work, we combine numerical PIC simulations with a simple analytical model to determine physical parameters of the beam and plasma, so that (i) the problem can be treated with reasonable computational power, and (ii) the physical picture is adequate to realistic pair beams. The physical picture is determined by several aspects: (i) the ratio of the energy densities of the beam and background plasma, (ii) instabilities and their growth rates, and (iii) non-linear damping of plasma waves. Here, we are concerned only with the first two subjects, and the non-linear effects \citep{Lazar03,Liu11} will be analyzed in future papers. 
Such treatment is possible in the linear stage which we are interested in here.

The created pairs are subject to the ICS and a full electromagnetic cascade that modifies their parameters. However, the goal of the present paper is to explore the potential dominance of plasma effects on the beam evolution. Therefore, we consider a pair beam created only by the initial TeV gamma-ray emission neglecting ICS. In this case, the typical parameters of the created beams depend on the distance from a blazar, and they are $\langle \Gamma \rangle=10^5$, $\Gamma=10^3-10^8$, $n_b=10^{-25}-10^{-19}$ cm$^{-3}$, $\Delta\theta\approx1/\langle \Gamma\rangle\approx10^{-5}$ ($\Delta\theta$ is the angular spread), whereas typical parameters of the IGM are $T=10^4-10^7$ K, $n=10^{-7}$ cm$^{-3}$ \citep{Broderick12,RS12,Sironi14,Miniati13}. Thus, the energy density ratio is $\epsilon=n_b\langle\Gamma\rangle m_e c^2/ (n k_B T)\approx  10^{-10}-10^{-1}$ ($k_B$ is the Boltzmann constant, $m_e$ is the electron mass) indicating that the pair beam cannot considerably heat the IGM plasma. This point was realized by \citet{Kempf16} who conducted simulations for $\epsilon=0.1$. The parameters of the simulations by \citet{Sironi14} are $\alpha=n_b/n\approx 10^{-2}$, $\Gamma\approx10^2$, and $k_BT/(m_e c^2)\approx 10^{-8}$, providing with $\epsilon\approx 10^8$, a parameter regime that is not relevant for realistic pair beams. Moreover, such a high energy-density ratio causes anisotropic plasma heating that can eventually drive the Weibel instability as it will be shown below. Note that Kempf and Sironi have studied a beam distribution with $\Delta\Gamma\ll \langle\Gamma\rangle$. We will also investigate this case in the present work, whereas a realistic distribution with $\Delta\Gamma\gg\langle\Gamma\rangle$ will be studied in a separate paper. 

The pair beam can induce two unstable modes: electrostatic and electromagnetic. The growth rate of these instabilities sensitively depends on the momentum spread of the beam. If the momentum spread is small enough, then the instabilities evolve in the so-called reactive regime. In this case, the beam can be mathematically treated as a delta function \citep{RS12} and the growth rates of the electrostatic and electromagnetic instabilities are maximal perpendicular to the direction of the beam propagation \citep{Godfrey75}. As the momentum spread increases, the electromagnetic instability becomes stabilized \citep{Bret05}, while the maximum growth rate of the electrostatic mode shifts to the direction parallel to the beam propagation \citep{Breizman90}. This is the so-called kinetic regime. \citet{Miniati13} have argued that the momentum spread of the realistic pair beam drastically reduces the growth rate of the electrostatic instability. Later, \citet{RS13} have disputed this statement. \citet{Sironi14} have demonstrated that the maximum growth rate occurs in the direction almost parallel to the beam (contrary to the reactive regime, when the maximum growth rate occurs in quasi-perpendicular direction to the beam). But \citet{RS13} have assumed the parallel direction of the wave vector from the very beginning. In this case, the electrostatic growth rate, indeed, only weakly depends on the beam temperature \citep{Bret05}. Thus, we can conclude that the electrostatic instability for a blazar-induced beam evolves in the kinetic regime at all angles with the maximum growth rate parallel to the beam propagation. It can be shown (see below) that for the beam parameters used by \citet{Kempf16} and \citet{Sironi14} the electrostatic instability has evolved in the reactive regime. Thus, an adequate behaviour of the instability has not been simulated before.  

So far we have discussed only the electrostatic instability. Usually, the electromagnetic (Weibel) instability can be neglected due to its smaller growth rate, but that is not always the case. If we compare the growth rate of the parallel electrostatic instability with the maximum Weibel growth rate $\gamma_W\approx (V_b/c)(\alpha/\Gamma)^{1/2}$, then $\gamma_W/\gamma_{react,\parallel}\approx\alpha^{1/6}\Gamma^{1/2}=0.3-1$ for $\alpha=10^{-18}-10^{-15}$. Thus, the Weibel instability can be potentially competitive with the electrostatic one. Note that we have used $\gamma_W$ assuming that the beam does not have any momentum spread, and the situation can be different for a beam with a finite temperature. \citet{Bret05} have shown that the Weibel instability is strongly suppressed by the non-relativistic perpendicular temperature of the beam. In this work, we will investigate the case of a relativistic temperature and demonstrate that the Weibel instability is suppressed in the case of a realistic blazar-induced beam. Additionally, we will demonstrate that for other conditions (relevant for PIC simulations) this mode can grow. 

Consequently, three criteria for a physically relevant simulation setup can be specified: (i) the energy density ratio, $\epsilon$, must be much smaller than unity, (ii) the beam temperature must be high enough, so that the parallel electrostatic instability evolves in the kinetic regime at all angles, and (iii) the Weibel instability must be suppressed. The goal of the current work is to find parameters satisfying all these requirements and to model them using PIC simulations. 

In Sec. \ref{Analyt}, we develop a simple analytical model of plasma instabilities. In Sec. \ref{Kin}, we evaluate a condition for the parallel electrostatic instability to be in the kinetic regime. In Sec. \ref{Param}, we discuss our choice of physical parameters for PIC simulations. Sec. \ref{Simul} presents simulation results and their discussion. The final summary is given in Sec. \ref{Summary}.


\section{Analytical model}\label{Analyt}

We already noted that the electrostatic instability evolves in the kinetic regime and has its maximum growth rate in the direction almost parallel to the beam propagation. At the same time, \citet{RS13} demonstrated that the growth rate of the parallel electrostatic instability very weakly depends on the momentum spread of the beam. Therefore, we can use the well-known growth rate of the two-stream instability for a cold plasma,

\be
\gamma_\mathrm{TS}= \frac{3^{1/2}}{2^{4/3}}\omega_p\alpha^{1/3}\Gamma^{-1}.
\label{an1}
\ee

Thus, we need to investigate only the electromagnetic Weibel instability. It should be noted that the most unstable wave vector of the Weibel mode can be in transverse direction  to the beam \citep{Califano98} as well as in the oblique direction \citep{Bret10}. Moreover, the work by \citet{Bret10} shows that for dilute beams the maximum growth rates of Weibel mode in the transverse and oblique directions can differ by a factor 2. Therefore, to make a rough estimation, we will study the Weibel instability only for wave vectors perpendicular to the beam. The PIC simulations described in the next section should include oblique modes as well. To derive analytical results, the beam-plasma system is modeled by a waterbag distribution \citep{Bret05,Yoon87}. Then, the distributions of the beam and the plasma, respectively, are

\begin{multline}
f_b({\bf p})= {n_b\over 4p_{\perp,b}^2 (p_{\parallel,b}^{+}-p_{\parallel,b}^{-})} 
\l[ \theta\l(p_z+p_{\perp,b}\r)-\theta\l(p_z-p_{\perp,b}\r) \r] \\ \times \l[ \theta\l(p_y+p_{\perp,b}\r)-\theta\l(p_y-p_{\perp,b}\r) \r] \l[ \theta\l(p_x-p_{\parallel,b}^{-}\r)-\theta\l(p_x-p_{\parallel,b}^{+}\r) \r] ,
\label{an2}
\end{multline}

\begin{multline}
f_p({\bf p})= {n\over 8p_{\perp,p}^2 p_{\parallel,p} }  
\l[ \theta\l(p_z+p_{\perp,p}\r)-\theta\l(p_z-p_{\perp,p}\r) \r] \\ \times \l[ \theta\l(p_y+p_{\perp,p}\r)-\theta\l(p_y-p_{\perp,p}\r) \r] \l[ \theta\l(p_x+p_{\parallel,p}\r)-\theta\l(p_x-p_{\parallel,p}\r) \r] ,
\label{an3}
\end{multline}
where $p_{\parallel,b}^{\pm}=p_0\pm p_{\parallel,b}$; $p_0$ is the beam drift momentum; $p_{\parallel,b}$ and $p_{\perp,b}$, respectively, the parallel and perpendicular momentum spreads of the beam; $p_{\parallel,p}$ and $p_{\perp,p}$, respectively, the parallel and perpendicular momentum spreads of the background plasma; $\theta(x)$, the Heaviside step function. The beam and background plasma are assumed to be homogeneous with number densities, accordingly, $n_b$ and $n$. It is useful to consider separately two cases: 
(i) $p_{\perp,b}=p_{\parallel,p}=p_{\perp,p}=0$ and (ii) $p_{\parallel,b}=0$.

\subsection{Case $p_{\perp,b}=p_{\parallel,p}=p_{\perp,p}=0$}\label{Case1}

We derive the dispersion equation for this case in Appendix A. It reads

\begin{multline}
\l[ 1-{\omega_p^2\over\omega} -{\omega_b^2\over\omega^2}U_1 \r]
\l\{  1- \l(kc\over\omega \r)^2 - {\omega_p^2\over\omega^2}- \r. \\ \l. -{\omega_b^2\over\omega^2} \l[ \l(kc\over\omega \r)^2 U_1   +  \l( 1- \l(kc\over\omega \r)^2\r)U_2  \r] \r\}- \\ - \l({\omega_b^2\over\omega^2}{kc\over\omega}U_3 \r)^2=0.
\label{c1} 
\end{multline}

Taking the limiting case $p_{\parallel,b}\ll p_0$, Eq. (\ref{c1}) provides the classical text book result \citep{Breizman90}

\begin{multline}
\l[ 1-{\omega_p^2\over\omega} -{\omega_b^2\over\Gamma\omega^2} \r]
\l\{  1- \l(kc\over\omega \r)^2 - {\omega_p^2\over\omega^2}-   {\omega_b^2\over\Gamma^3\omega^2}- \r. \\ \l. - {\omega_b^2\over\Gamma\omega^2}\l(kV_o\over\omega \r)^2     \r\}-  \l({\omega_b^2\over\Gamma\omega^2}{kc\over\omega} \r)^2=0,
\label{c2} 
\end{multline}
where $V_0=p_0/(m_e\Gamma)$. Eq. (\ref{c2}) predicts an instability with growth rate \citep{Godfrey75}:

\be
\gamma=\omega_p\beta_0 \l(\alpha\over\Gamma \r)^{1/2}{kc\over \l(\omega_p^2+ (kc)^2 \r)^{1/2}}.
\label{c3}
\ee

Now, we will show that the solution (\ref{c3}) is only slightly different for a large parallel momentum spread $p_{\parallel,b}\gg p_0$. Assuming $p_{\parallel,b}^{+}\gg mc$ and $p_{\parallel,b}^{-}\ll mc$, we obtain

\be
U_1\approx U_3\approx {m_ec\over p_{\parallel,b}^{+}} \ln {p_{\parallel,b}^{+}\over m_ec},
\label{c4}
\ee

\be
U_2\approx {m_ec\over p_{\parallel,b}^{+}}.
\label{c5}
\ee

Neglecting unity in each bracket in Eq. (\ref{c1}) results in the solution

\be
\gamma= \omega_p \l[ \alpha {m_e c\over p_{\parallel,b}^{+}}\ln {p_{\parallel,b}^{+}\over m_ec} \r]^{1/2}{kc\over \l(\omega_p^2+ (kc)^2 \r)^{1/2}}.
\label{c6}
\ee

It is well seen from Eq. (\ref{c6}) that even for $p_{\parallel,b}^{+}=10^2 p_0$, the difference between the solutions (\ref{c5}) and (\ref{c6}) is only a factor of $0.4$. Thus, we can neglect the parallel momentum dispersion of the beam and use $p_{\parallel,b}=0$.

\subsection{Case $p_{\parallel,b}=0$}

The dispersion equation for $p_{\parallel,b}=0$ is derived in Appendix B and has the following form 

\begin{multline}
\l[ 1 - {\omega_{p,p}^2\over\omega^2}- {\omega_{p,e}^2 \over \omega^2- (kv_{\perp,p})^2} - {\omega_{b,\perp}^2\over \omega^2 - (ku)^2} \r] \times \\ \times 
\l[ 1 - {(kc)^2+\omega_p^2 + \omega_{b,\parallel}^2 \over \omega^2} -  {(kv_{\parallel,p})^2 \omega_{p,e}^2\over 3\omega^2( \omega^2 - (kv_{\perp,p})^2 )   } - \r. \\ \l. 
- {(kV_0)^2\omega_{b,\perp}^2 \over \omega^2 (\omega^2 - (ku)^2)}
\r] - \l[ {kV_0\omega_{b,\perp}^2 \over\omega(\omega^2- (ku)^2 )} \r]^2=0,
\label{d1} 
\end{multline}
where $\omega_{b,\perp}^2=\omega_b^2/\Gamma$, $\omega_{b,\parallel}^2=\omega_b^2/\Gamma^3$. In principle, one can analyze Eq. (\ref{d1}) analytically, but it is more useful and easier to treat two limiting cases of the cold background plasma and the cold beam.

\subsubsection{Cold background plasma $v_{\parallel,p}=v_{\perp,p}=0$}

Neglecting unity in each bracket in Eq. (\ref{d1}), we obtain that for $p_{\perp,b}\geq p_0(\alpha/\Gamma)^{1/2}$ the solution is purely real (no instability can arise), whereas for $p_{\perp,b}< p_0(\alpha/\Gamma)^{1/2}$ the Weibel mode is unstable for $k<(\omega_p/c)[(\alpha/\Gamma)(p_0/p_{\perp,b})^2-1]^{1/2}$ with growth rate

\be
\gamma= \l( \omega_p^2\beta_0^2{\alpha\over\Gamma}{(kc)^2 \over (kc)^2+\omega_p^2 } - (ku)^2 \r)^{1/2} .
\label{d2}
\ee

Let us now assume that the beam obeys a relativistic Maxwellian distribution: 
\be
f({\bf p})= {\mu \over 4\pi (m_ec)^3\Gamma^2 K_2(\mu/\Gamma)}e^{-\mu \l[ \l( 1+{p^2\over(m_ec)^2} \r)^{1/2}- \beta_0 {p_x\over m_ec}  \r]},
\label{k3}
\ee
where $\mu=\Gamma \mu_R =\Gamma m_ec^2/(k_BT_b)$, $\beta_0=V_0/c$. Here, $T_b$ is the temperature of the beam in its rest frame. Then we can evaluate $\Delta p_\perp$ (see Appendix C) and write the condition for the Weibel mode stability as
\be
\alpha\leq \alpha_W= {2-\pi/2 \over \Gamma \mu_R}=\left(2-\frac{\pi}{2}\right) \frac{k_BT_b}{\Gamma m_ec^2}.
\label{d2.1}
\ee

In the simulations by \citet{Sironi14}, magnetic-field fluctuations grew at early times due to the Weibel instability driven by the beam, because $\alpha=10^{-2}$ and $\alpha_W<5\times10^{-4}$ led to condition (\ref{d2.1}) not being fulfilled. But in the simulations by \citet{Kempf16}, the Weibel mode was suppressed, because $\alpha=2\times10^{-6}<\alpha_W=10^{-5}$. For a realistic blazar-induced beam, the Weibel instability is also suppressed, since $k_BT_b\approx m_ec^2$ and $\alpha \ll 1/<\Gamma>$.

\subsubsection{Cold beam $p_{\perp,b}=0$}

Again neglecting unity in Eq. (\ref{d1}), we can approximate it as

\be
E\omega^4 + F\omega^2 +G = 0.
\label{d3}
\ee
where

\be
E= (kc)^2+\omega_p^2 , 
\label{d4}
\ee

\be
F= \l(kV_0\omega_{b,\perp}\r)^2+ {1\over3}(kv_{\parallel,p}\omega_{p,e})^2  - (kv_{\perp,p})^2((kc)^2+\omega_p^2) ,
\label{d5}
\ee

\be
G= - \l(kV_0\r)^2(kv_{\perp,p})^2\omega_{b,\perp}^2.
\label{d6}
\ee

The growth rate reads

\be
\gamma= \l[F+ (F^2-4EG)^{1/2} \over 2E  \r]^{1/2}\approx \l[ {F+|F|\over2E}- {G\over|F|}\r]^{1/2}. 
\label{d7}
\ee

If $F>0$, Eq. (\ref{d7}) describes the classical Weibel instability \citep{Bret05} with growth rate

\be
\gamma\approx \l( F\over E\r)^{1/2} \approx \l( (kv_{\parallel,p}\omega_{p,e})^2/3  - (kv_{\perp,p})^2((kc)^2+\omega_p^2) \over (kc)^2+\omega_p^2  \r)^{1/2}.
\label{d8}
\ee

\begin{figure}
\includegraphics[width=90mm]{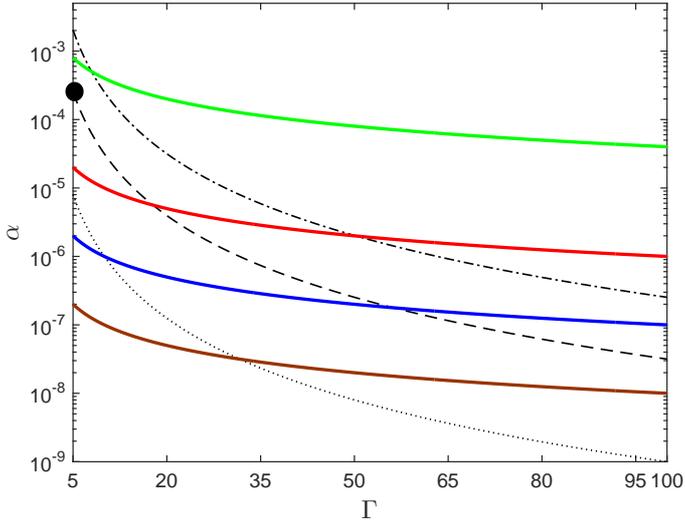}
\caption{Function $\alpha_{kin}(\Gamma)$: dashed dotted black line $\mu_R=2.5$; dashed black line $\mu_R=10$; dotted black line $\mu_R=100$. Function $\alpha_{\epsilon}(\Gamma)$: green line $k_BT_p/(m_ec^2)=4\times10^{-3}$; red line $k_BT_p/(m_ec^2)=10^{-4}$; blue line $k_BT_p/(m_ec^2)=10^{-5}$; brown line $k_BT_p/(m_ec^2)=10^{-6}$. The black point illustrates the parameters chosen for simulation run 1 that satisfies all criteria. }
\label{Parameter}
\end{figure}

Due to $v_{\perp,p}\neq0$, the instability is stabilized at large wave vectors, but at small $k$ the plasma is unstable for $(v_{\parallel,p}/v_{\perp,p})^2>3$. These conditions were fulfilled in the simulations by \citet{Sironi14}, where there was a growth of the magnetic-field fluctuations at later time around $\omega_{p,e}t\approx 10^4$. In the opposite case $F<0$ and assuming that $v_{\perp,p}$ is large enough, Eq. (\ref{d7}) reduces to Eq. (\ref{c3}). 

\section{Condition for the kinetic regime}\label{Kin}

The parallel electrostatic instability evolves in the kinetic regime \citep{Breizman90}, if

\be
\l| v_{\parallel,b} \over c \r|\gg \alpha^{1/3}\Gamma^{-1} 
\label{k1}
\ee
which can be re-written as
\be
\alpha\ll \alpha_{kin}= \l(\Gamma \l| \frac{v_{\parallel,b}}{c} \r| \r)^3.
\label{k2}
\ee

An analytical expression for $\alpha_{kin}$ is derived in Appendix C, and its functional behavior is illustrated in Fig. \ref{Parameter}. 

For the simulation parameters used by \citet{Kempf16}, $\mu_R=5\times10^3 $ ($T_b=10^6$ K), $\Gamma=10$, and $\alpha=2.5\times10^{-6}$, we obtain $\alpha_{kin}\approx2.8\times 10^{-9}$ and Eq. (\ref{k2}) is not fulfilled. 
For the work of \citet{Sironi14}, $\Gamma=300$, $\alpha=10^{-2}$, $\mu>3$, it results in $\alpha_{kin}\approx 7.1\times10^{-9}$, and Eq. (\ref{k2}) is not satisfied again. Hence, both \citet{Kempf16} and \citet{Sironi14} did not simulate the electrostatic instability in the appropriate kinematic regime of pair cascades from AGN.

\begin{figure}
\includegraphics[width=90mm]{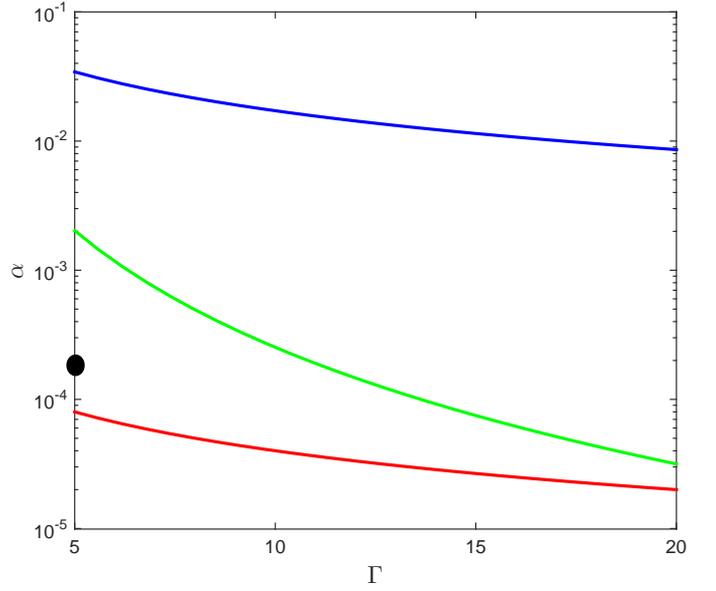}
\caption{Dependence of all three constraints on the beam Lorentz factor. The green line represents $\alpha_{kin}$, the red line $\alpha_\epsilon$, and the blue line $\alpha_W$. $\mu_R=2.5$, $k_BT_p/(m_ec^2)=4\times10^{-4}$. The black dot indicates parameter values of run 2.}
\label{ParamCase1}
\end{figure}

\begin{figure}
\includegraphics[width=90mm]{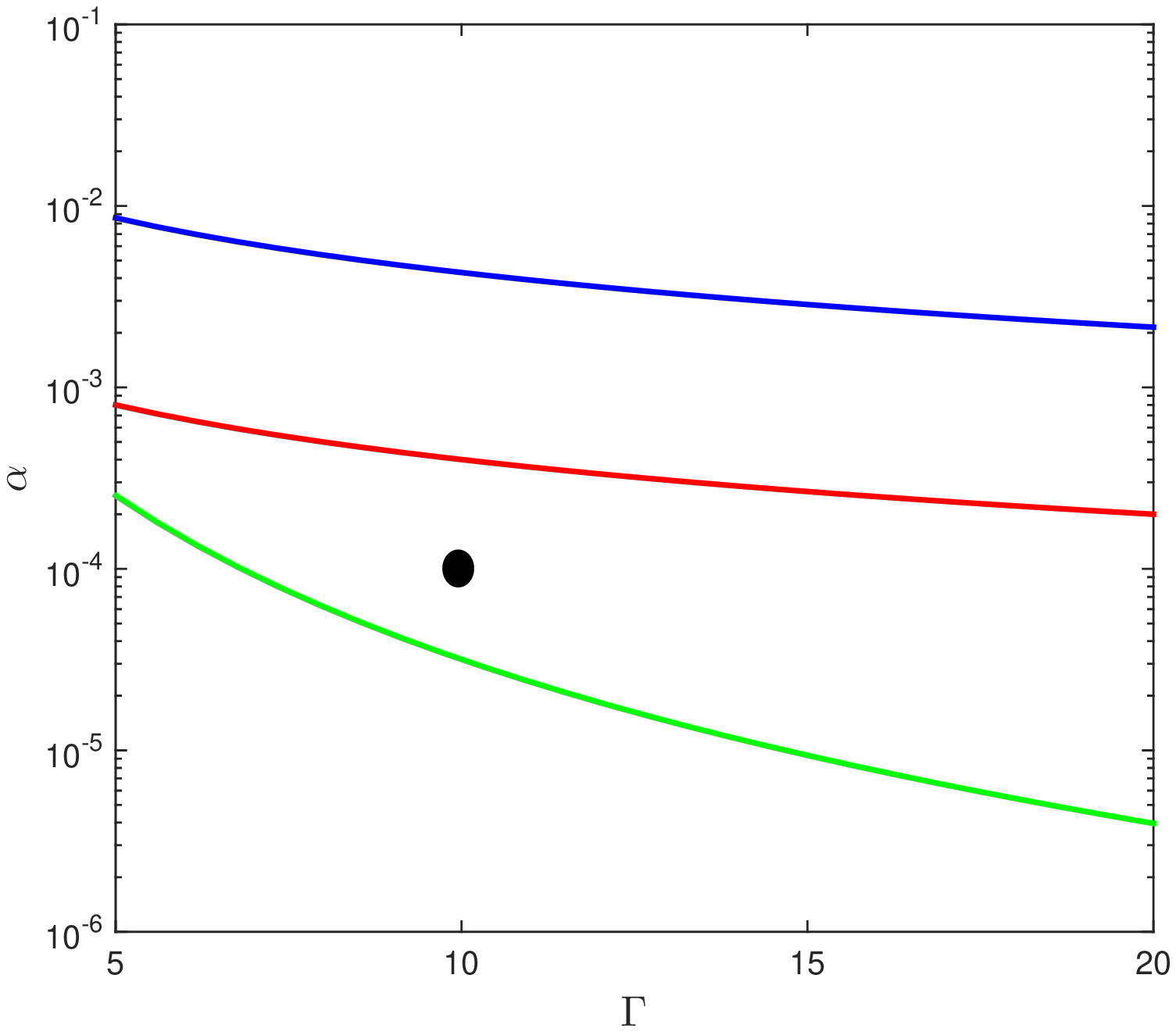}
\caption{Same as Fig. \ref{ParamCase1}, but for $\mu_R=10$, $k_BT_p/(m_ec^2)=4\times10^{-3}$. The black dot indicates parameter values of run 3.}
\label{ParamCase2}
\end{figure}

\begin{figure}
\includegraphics[width=90mm]{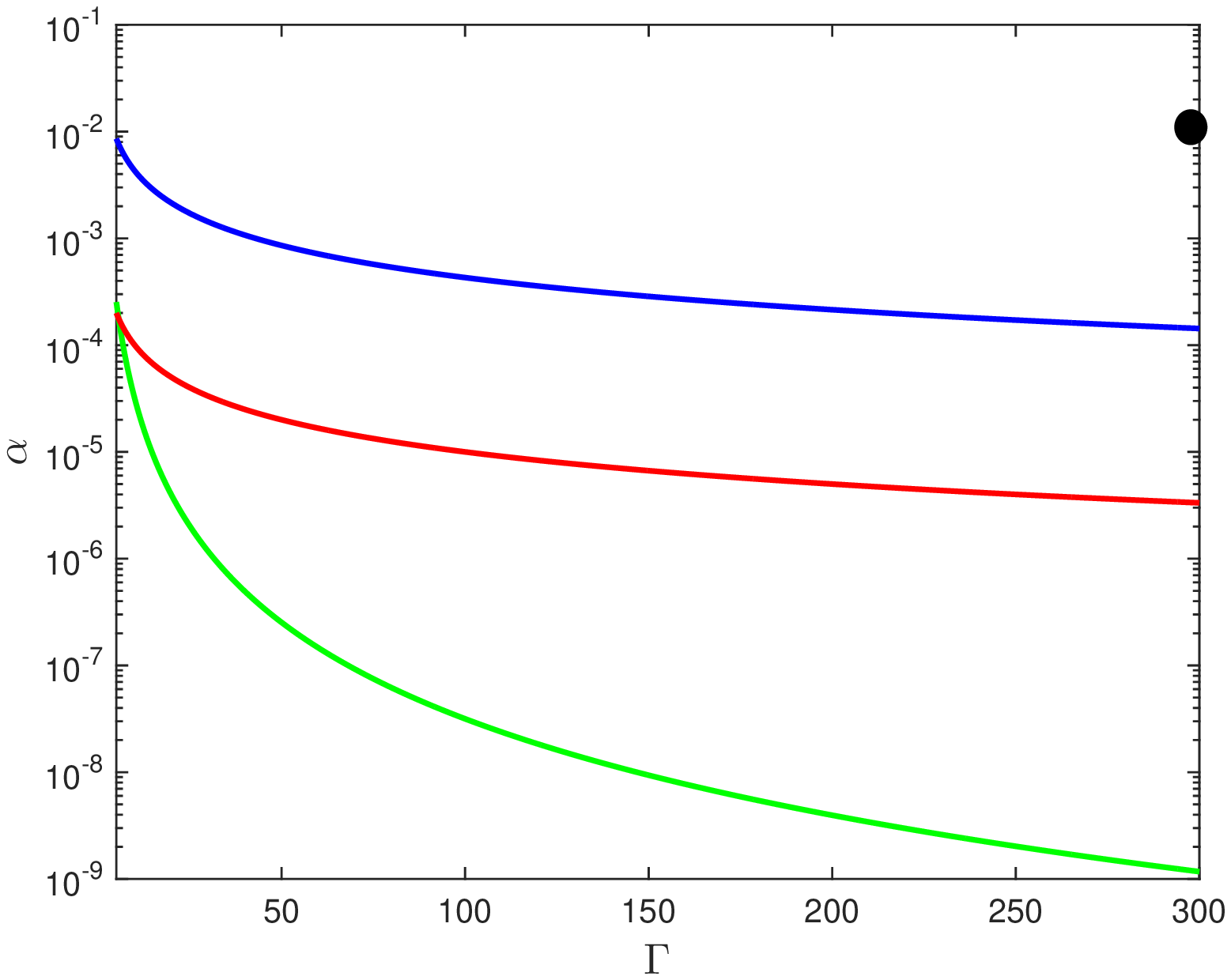}
\caption{Same as Fig. \ref{ParamCase1}, but for $\mu_R=10$, $k_BT_p/(m_ec^2)=10^{-3}$. The black dot indicates parameter values of run 4.}
\label{ParamCase3}
\end{figure}

\section{Choice of parameters for PIC simulations}\label{Param}

In the introduction, we have specified three criteria for a physically relevant setup for the beam-plasma system. First, the energy density ratio must satisfy $\epsilon=\alpha\Gamma m_ec^2/(k_B T_p)\ll1$ yielding 
\be
\alpha\ll \alpha_{\epsilon}(\Gamma)= {k_BT_p\over \Gamma m_ec^2}\approx {10^{-6}-10^{-3}\over \Gamma}.
\label{p1}
\ee
The behavior of $\alpha_{\epsilon}(\Gamma)$ is shown in Figures~\ref{ParamCase1}, \ref{ParamCase2}, and \ref{ParamCase3} as red line.
Second, the electrostatic instability should develop in the kinetic regime at all angles which is determined by Eq. (\ref{k2}), for which we indicate $\alpha_{kin}$ by the green line in the figures.
Lastly, the Weibel mode must be stable which requires satisfying Eq. (\ref{d2.1}). Eq. (\ref{d2.1}) is automatically fulfilled due to $\alpha_{kin}\ll\alpha_W$ for $\Gamma>1$ and $\mu_R>1$. Fig. \ref{Parameter} compares the functions $\alpha_{kin}(\Gamma)$ and $\alpha_{\epsilon}(\Gamma)$. To satisfy Eqs. (\ref{k2}) and (\ref{p1}) for given values of $\Gamma$ and $\mu$, the value of $\alpha$ must be below both curves $\alpha_{kin}(\Gamma)$ and $\alpha_{\epsilon}(\Gamma)$. We defined a simulation setup, henceforth referred to as run 1, that would satisfy all criteria. The main parameter values are $\Gamma=5$ and $\alpha=2\times10^{-4}$, and it is indicated in Figure~\ref{Parameter} by a black dot. 

In addition, we have specified three other setups (runs 2-4) that are listed in 
Table \ref{Table1}. The goal of these tests is to determine the impact of a violation of one of the criteria on the beam-plasma evolution. For run 2, the energy density ratio $\epsilon=2.5$ is higher than unity, and one might expect a strong heating of the background plasma and subsequently the development of other instabilities. Run 3 considers the evolution of the electrostatic instability in the reactive regime ($\alpha>\alpha_{kin}$), and beam energy losses are expected to be larger. Finally, all the conditions are violated for run 4. The values of $(\alpha;\Gamma)$ for runs 2-4 are demonstrated by the black dots in Figs. \ref{ParamCase1}-\ref{ParamCase3}, respectively.

\section{The simulation code}\label{SimuC1}
For the simulation purposes we use EPOCH 2D, a multi-dimensional, fully electromagnetic, relativistic particle-in-cell code developed by the Collaborative Computational Plasma
Physics (CCPP) consortium and funded
by the Engineering and Physical Sciences Research Council
(EPSRC). PIC codes solve Maxwell’s equations on a numerical (Eulerian) grid and follow charged computational particles (CP) as they move under the influence of the electromagnetic field and provide charge and current density \citep{Dawson83,Birdsall04}.

The relevant equations are
\be 
\mu_0\epsilon_0  \frac{\partial E(x,t)} {\partial t} ={\nabla}\times B(x,t) -\mu_0 J(x,t)
\label{p22}
\ee
and 
\be 
\frac{\partial B(x,t)} {\partial t} =-{\nabla}\times E(x,t) \ ,
\label{p23}
\ee
where the current density, $J(x,t)$, is computed using the algorithm of \citet{Villasenor92}.
Collisionless plasma is set up with a Maxwellian velocity distribution. For each CP the field pusher solves the relativistic equation of motion with a numerical approximation of Lorentz force equation. EPOCH is  a refined version of the basic explicit PIC algorithm with higher-order weights and interpolation schemes \citep{Arber15}. 
Note that the 2D model can break down on the non-linear evolution stage, when 3D mode coupling becomes important \citep{Lazar03,Liu11}. As the electrostatic mode involves a narrow resonance, its modeling in a PIC simulation requires a very good wavenumber resolution of the numerical grid \citep{2017arXiv170400014S}. This implies a large number of grid points in any direction which we can establish only in 2D. Waves of arbitrary orientation will be included, albeit with only one linear polarization, as is nonlinear wave coupling, provided it does not build on the polarization out of the simulation plane. In the current study we are mainly interested in the linear growth of the instabilities, and so we accept these limitations.

The simulation resolves the x–y plane with periodic boundary conditions. The simulation volume is filled with a beam of electrons and positrons and the background plasma of protons and electrons with real mass ratio. We performed a series of tests to verify the stability of the simulation against numerical artifacts. Of particular interest is avoiding artificial plasma heating arising from electric-field noise caused by the charge-density granularity in a particle simulation. We found that using 400 particles per cell and species is required to keep the plasma temperature as desired and the electric-field noise at a level significantly below the intensity of the electrostatic mode. The desired density ratio, $\alpha=n_b/n$, is established with numerical weights. The simulation box contains $1024 \times1024$ cells, each 1/8 of the skin length in size,  $ \lambda_{e}=\frac{c}{\omega_{pe}}=8\,\Delta_x$. The timestep is chosen to satisfy the CFL condition and to resolve the plasma frequency, $ \omega_{pe}=(n_{0}e^{2}/\epsilon_{0}m_{e})^{1/2}$.  

Table~\ref{Table1} lists the temperature of the IGM plasma, $T_p$, and of the beam in its rest frame, $T_b$. The IGM particles are initially at rest, while the beam is moving in x-direction with Lorentz factor $\Gamma_b$. For the IGM, EPOCH code generates a non relativistic thermal distribution using the method of \citet{BoxM58}. However, we implemented the algorithm of \citet{Zenitani15} to set up the relativistic Maxwellian distribution for the beam. 
For the graphical presentation 
we use the following normalization: distance and time
are normalized to $\frac{c}{\omega_{p,e}}$ and $\omega_{p,e}^{-1}$, and electric and magnetic fields are given in units of $\omega_{p,e}cm_e/e $ and $\omega_{p,e}m_e/e $, respectively. 

\begin{table}
\caption{Simulation parameters}
\centering
\label{Table1}
\begin{tabular}{|c|c|c|c|c|}    
\hline
run              & $\alpha$ & $T_b$ & $T_p$ & $\Gamma_b$ \\
\hline
 1 & 2.E-04  & 200 keV & 2 keV & 5   \\
 2            & 2.E-04  & 200 keV & 200 eV & 5   \\
 3           & 1.E-04  & 50 keV & 2 keV & 10  \\
 4     & 1.E-02  & 50 keV & 500 eV & 300   \\
\hline
\end{tabular}
\end{table}

In order to reduce the well-known PIC-code phenomena of self heating and statistical noise, all simulations are performed with a high number of CPs (400 particles per species), a 6th-order field particle pusher, and a triangular-shaped cloud (TSC) shape function, with the peak of the triangle located at the position of the pseudoparticle. 

\section{Discussion of simulation results}\label{Simul}
\subsection{Run 1}

As mentioned above, for run 1 all relevant criteria for the beam are fulfilled. First of all, the beam/plasma energy density ratio $\epsilon=0.5$ for run 1 is smaller than unity. Moreover, the beam is stable with respect to the Weibel instability, while the electrostatic mode grows as expected in the kinetic regime, i.e. at the parallel wave vector $k_{||}\approx\omega_p/c$ to the beam.

\begin{figure}
\includegraphics[width=90mm]{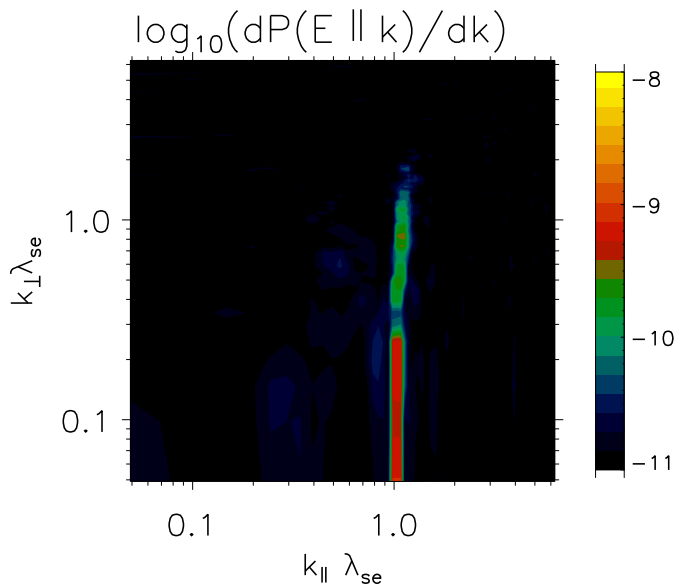}
\caption{Two-dimensional Fourier spectrum of $\mathbf{E} \parallel \mathbf{k}$ at $\omega_{pe}t=4222$ for run 1.}
\label{fig1}
\end{figure}

In Fig. \ref{fig1} we present the Fourier spectrum of the electric field, and it is evident that an electrostatic mode with $\mathbf{E} \parallel \mathbf{k}$ dominates with peak intensity for wave vectors roughly aligned with the beam direction. The linear growth rate of the electric field is about $\gamma\simeq 4\times10^{-4}\omega_{pe}$. The theoretically calculated maximum growth rate for parallel wave vectors is $5\times10^{-4}\omega_{pe}$ which approximately agrees with that derived numerically. 

Figs.~\ref{fig3} and \ref{fig4} demonstrate that after $16,237 \,\omega_{pe}^{-1}$, corresponding to about 8 growth times, the instability has saturated with negligible energy loss and heating of the beam. The latter is of interest because a widening of the lateral beam distribution would impose a temporal smearing of the ICS signal that would reduce the expected flux seen with Fermi-LAT. Our run 1 suggests that this effect is not efficient for realistic pair beams induced by gamma rays from AGN.

\begin{figure}
\includegraphics[width=90mm]{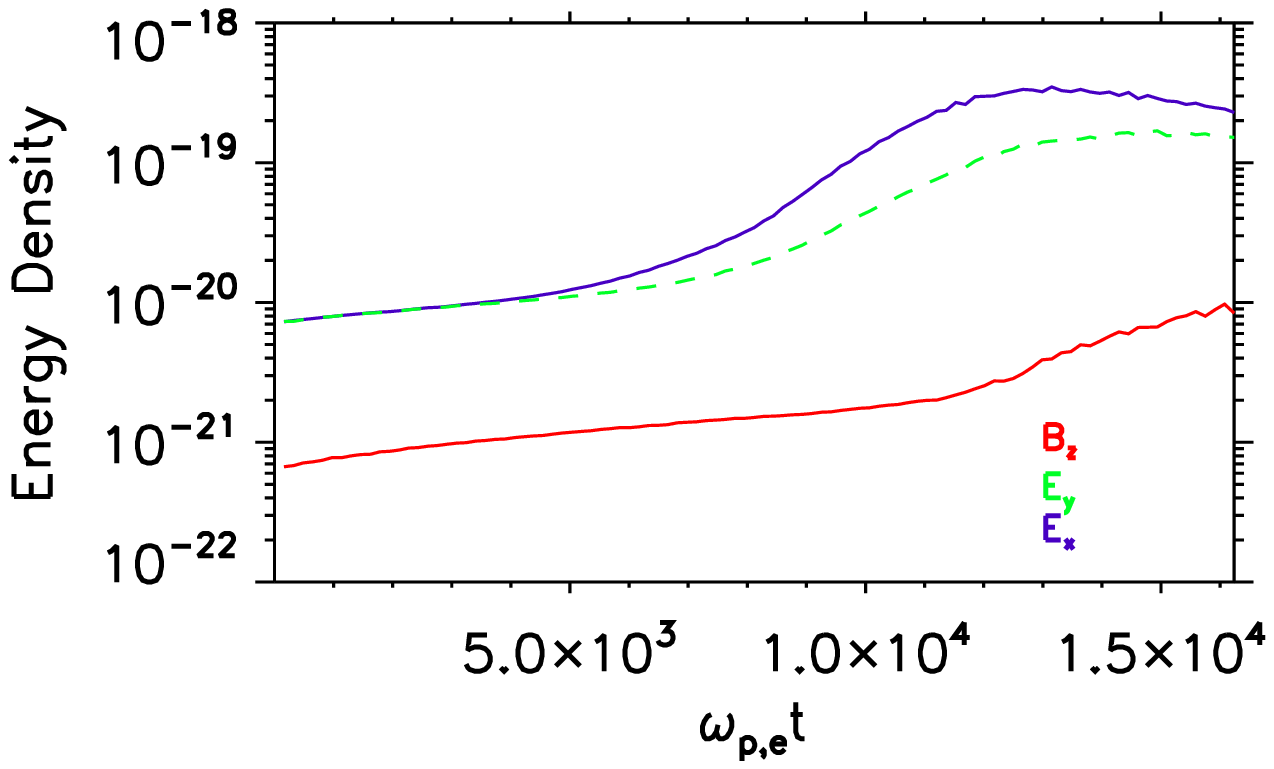}
\caption{Time evolution of the energy densities of electric and magnetic field, respectively, in SI units for run 1.}
\label{fig2}
\end{figure}

\begin{figure}
\includegraphics[width=90mm]{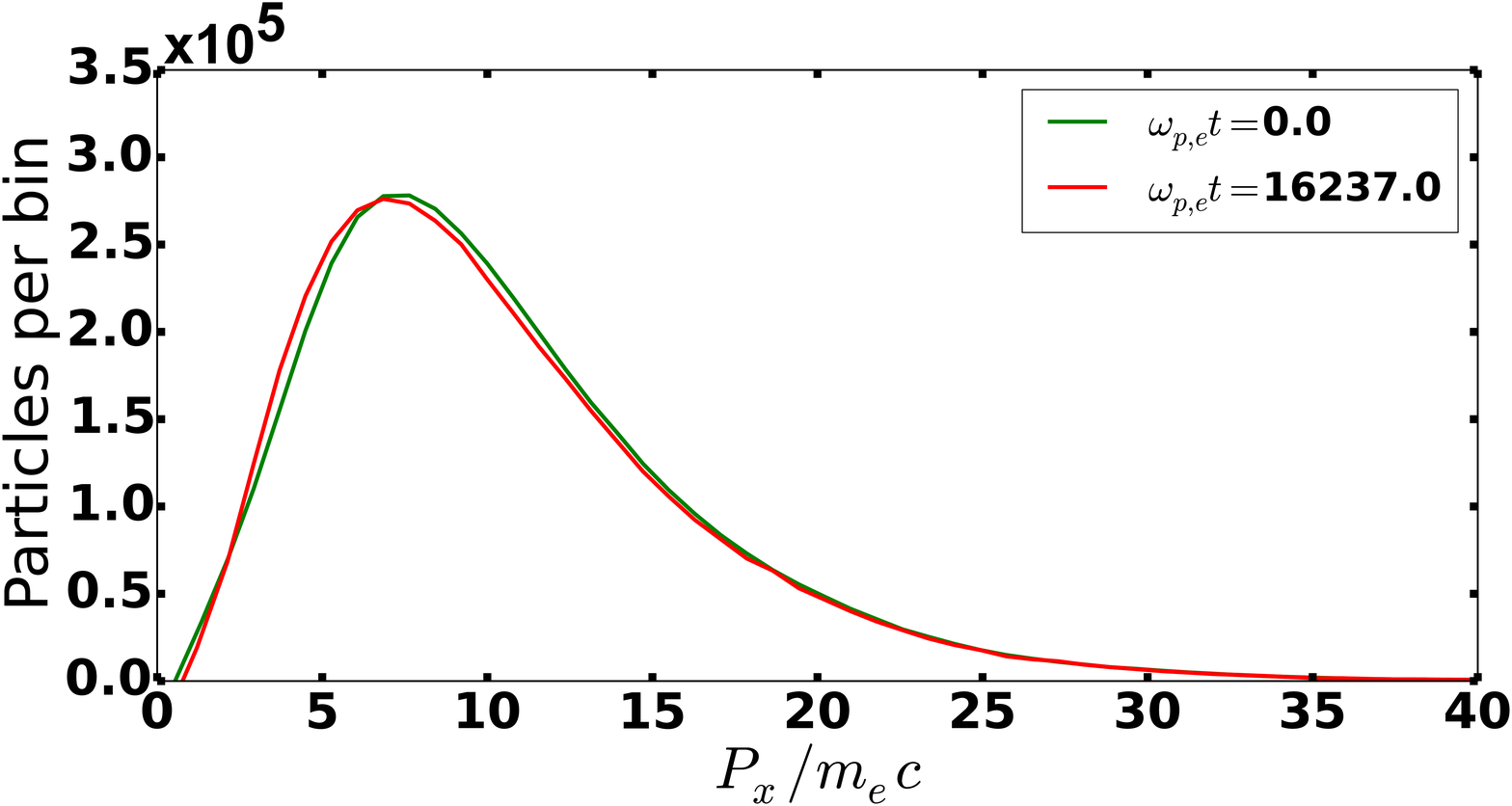}
\caption{Beam momentum distribution in $p_x$ for run 1  at two points in time.}
\label{fig3}
\end{figure}

\begin{figure}
\includegraphics[width=90mm]{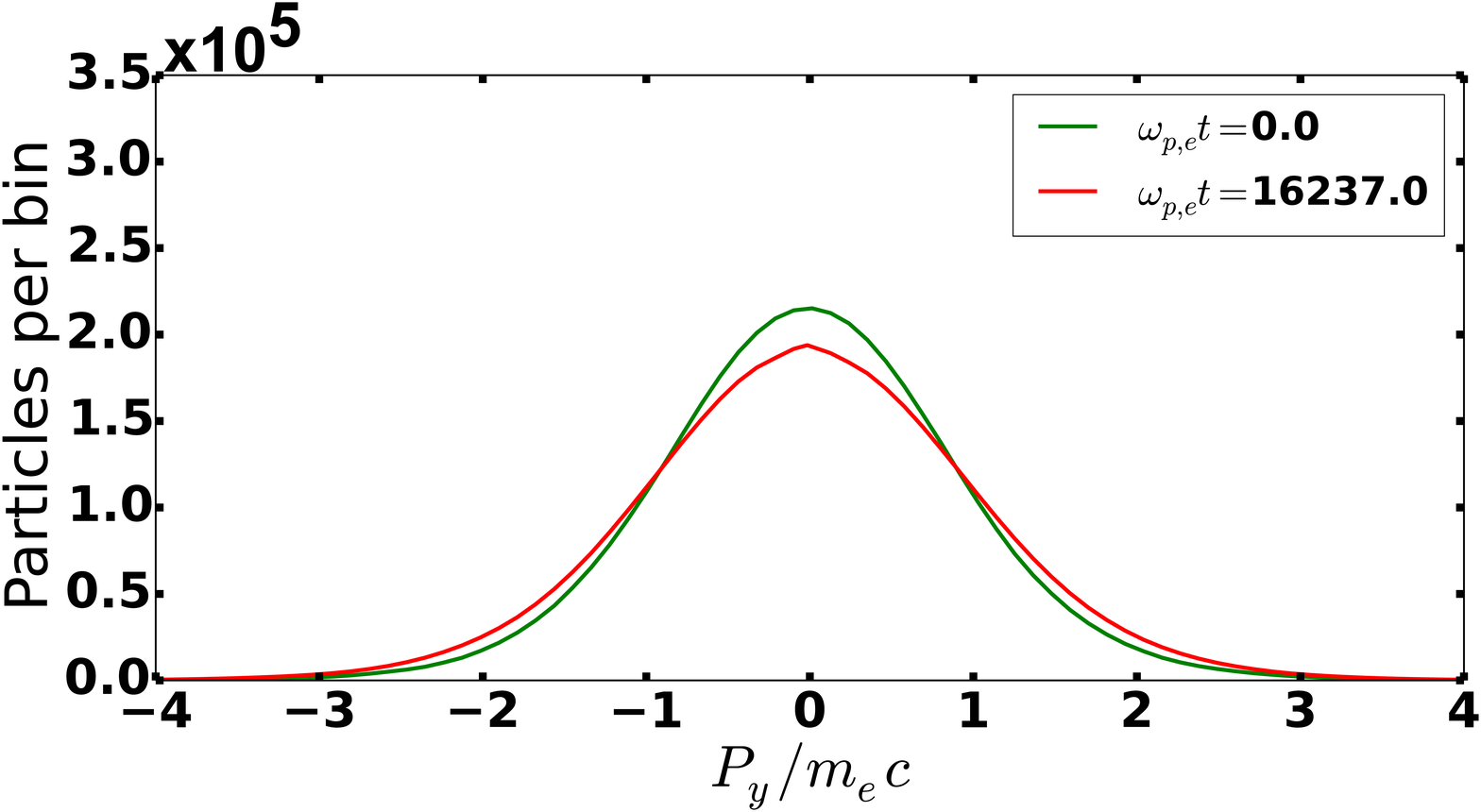}
\caption{Beam momentum distribution in $p_y$ for run 1 at two points in time.}
\label{fig4}
\end{figure}

Fig. \ref{fig2} illustrates the time evolution of the electric and magnetic field energy density. The electric field energy saturates after $\sim 7$ growth times. It is clear that the beam transferred only a tiny fraction ($\sim 10^{-4}$ \%) of its initial kinetic energy into the electromagnetic fields. Accordingly, the change of the beam distribution is also very small (see Figs. \ref{fig3}-\ref{fig4}). This development of the beam-plasma interaction is caused by the initial momentum spread of the beam. It was also found by \citet{Sironi14} that the beam momentum distribution does not relax to the plateau form when $\Delta p_{\perp,b}/m_e c\sim1$. The physical reason is that the electrostatic growth rate simply becomes much smaller than in the reactive regime. At the same time, the damping rates of the modulation instability and non-linear Landau damping depend on the resonant wave energy, and therefore they will stabilize the instability at smaller electric field energies. 

\subsection{Run 2}

\begin{figure}
\includegraphics[width=90mm]{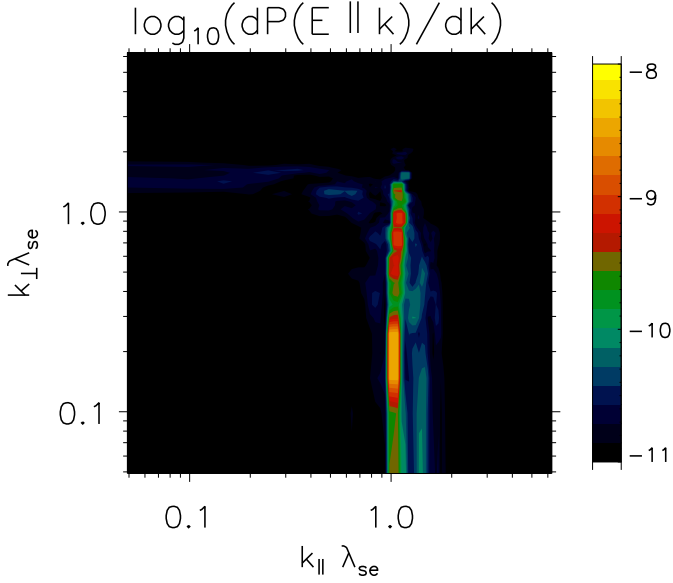}
\caption{Two-dimensional Fourier spectrum of $\mathbf{E} \parallel \mathbf{k}$ at $\omega_{pe}t=5036$ for run 2}
\label{fig5}
\end{figure}

\begin{figure}
\includegraphics[width=90mm]{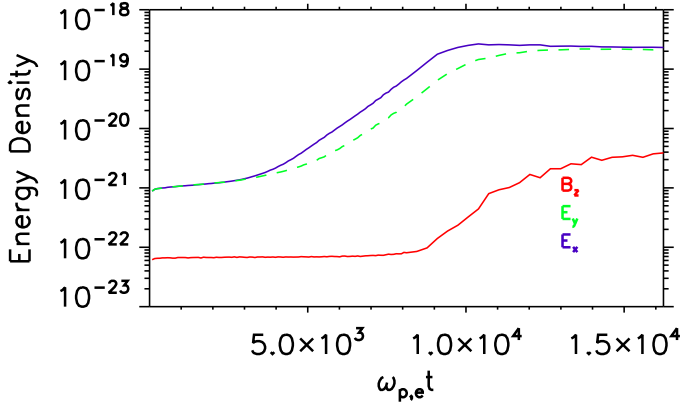}
\caption{Time evolution of the energy densities of electric and magnetic field, respectively, in SI units for run 2}
\label{fig6}
\end{figure}

In contrast to run 1, run 2 considers the beam/plasma energy-density ratio, $\epsilon=5$, greater than 1. The only parameter changed compared to run 1 is the plasma temperature that became by an order of magnitude smaller. Due to the fact that the beam parameters remained the same, the Weibel mode is still stable. The electrostatic instability also evolves in kinetic regime with a growth rate around $\simeq 5\times10^{-4}\omega_{pe}$, and the time evolution of the Fourier spectrum (shown in Fig. \ref{fig5} at $\omega_{pe}t=5036$) is consistent with the value.

Fig. \ref{fig6} shows that the electric field energy density saturates at nearly the same level as in run 1. Note that due to a smaller plasma temperature the initial electric noise level in run 2 is about by an order of magnitude smaller compared to run 1. Although the peak intensity of the electrostatic modes is now observed at a 10$^\circ$ angle to the beam direction, the distribution function again did not evolve appreciably, in particular not to a plateau distribution, and the beam experienced only a tiny energy loss or widening.

\subsection{Run 3}

\begin{figure}
\includegraphics[width=90mm]{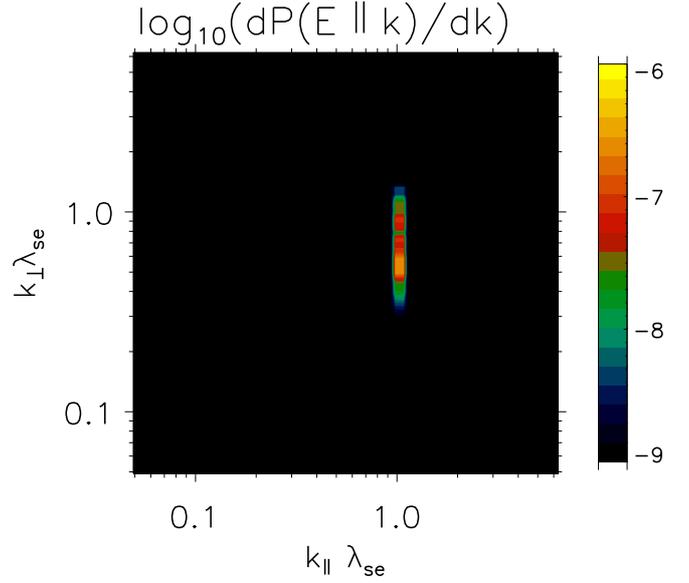}
\caption{Two-dimensional Fourier spectrum of $\mathbf{E} \parallel \mathbf{k}$ at $\omega_{pe}t=2448$ for run 3}
\label{fig7}
\end{figure}

\begin{figure}
\includegraphics[width=90mm]{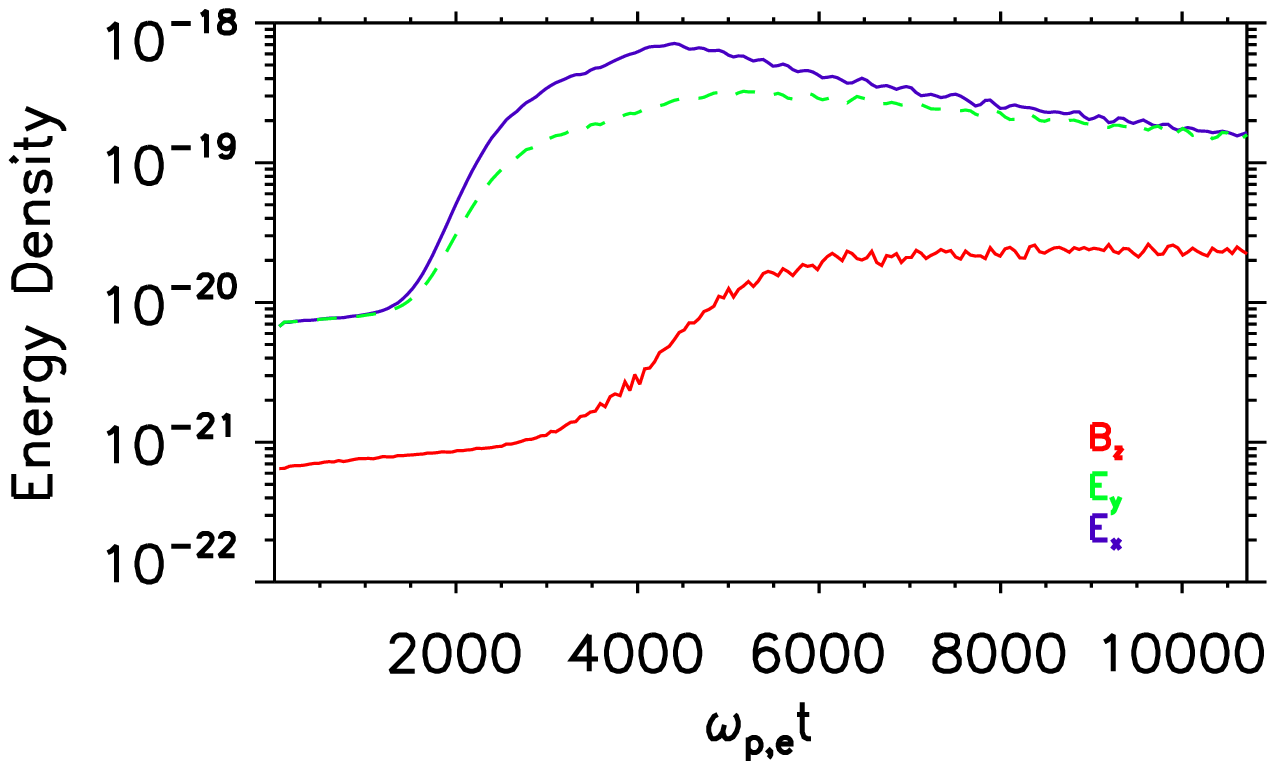}
\caption{Time evolution of the energy densities of electric and magnetic field, respectively, in SI units for run 3}
\label{fig8}
\end{figure}

\begin{figure}
\includegraphics[width=90mm]{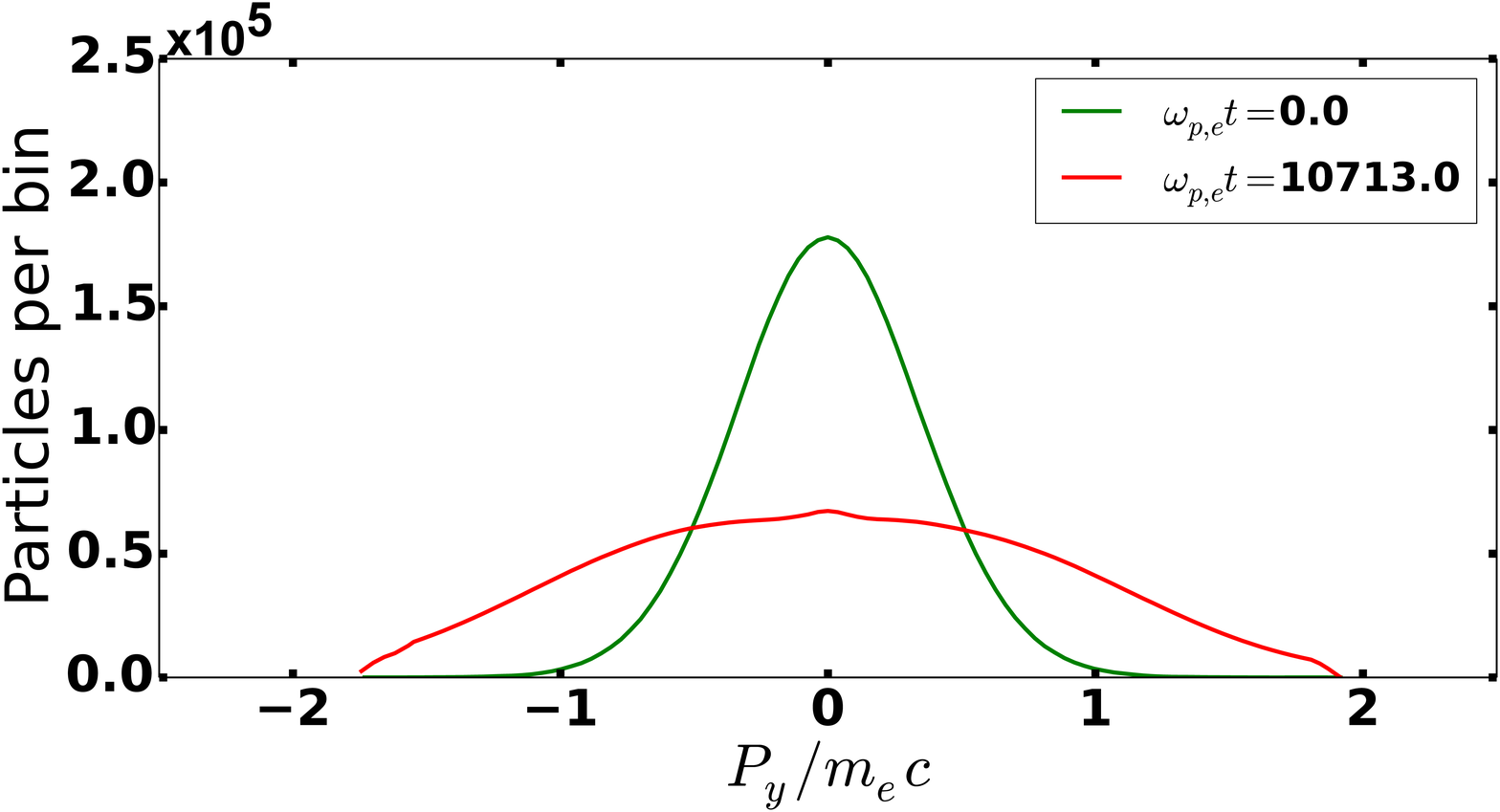}
\caption{Beam momentum distribution in $p_y$ for run 3 at two points in time.}
\label{fig9}
\end{figure}

\begin{figure}
\includegraphics[width=90mm]{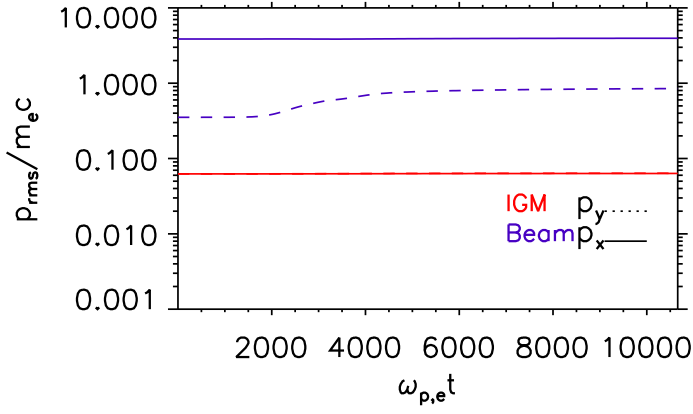}
\caption{Time evolution of the momentum spread of the beam, $p_\mathrm{rms}$, for run 3.}
\label{fig10}
\end{figure}

With run 3, we explore the reactive regime of the electrostatic mode in contrast to runs 1 and 2, where the instability was kinetic. To do this, we have reduced the temperature of the beam and increased its gamma factor. Now, the electrostatic instability grows at an oblique direction (at about 30$^\circ$) to the beam as is evident from the Fourier spectrum shown in Fig. \ref{fig7}. The growth rate for oblique propagation and the parameters of run 3 (assuming a cold beam \citep{Breizman90}) is
\be 
\gamma_\mathrm{TS}= \frac{3^{1/2}}{2^{4/3}}\omega_{pe}\l(\alpha\over\Gamma\r)^{1/3}\l({k_{||}^2\over k^2\gamma^2}  + {k_{\perp}^2\over k^2}\r)^{1/3}\simeq 9\times10^{-3}\omega_{pe} ,
\label{oblgr}
\ee
where the last equality applies for the parameters of run 3. The numerically determined growth rate is smaller than that by a factor 2-3. This difference may be explained by the fact that run 3 operates not very far from the condition $\alpha=\alpha_{kin}\l(\Gamma\r)$ (see Fig. \ref{ParamCase2}).

The instability growth rate of run 3 is larger by an order of magnitude compared to runs 1 and 2. Therefore, we can expect a more substantial modification of the beam. Although the electric-field energy density remains small as shown in Fig. \ref{fig8}, we observe in Fig. \ref{fig9} a significant transverse widening of the beam that is not seen in runs 1 and 2. Fig. \ref{fig10} indicates that the width of the perpendicular momentum distribution of the beam increased by a factor of 3.

\subsection{Run 4}

\begin{figure}
\includegraphics[width=90mm]{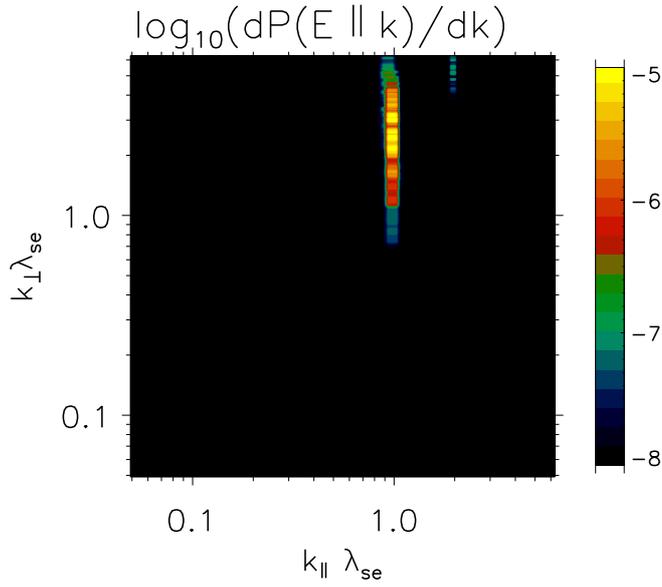}
\caption{Two-dimensional Fourier spectrum of $\mathbf{E} \parallel \mathbf{k}$ at $\omega_{pe}t=401$ for run 4.}
\label{fig11}
\end{figure}

\begin{figure}
\includegraphics[width=90mm]{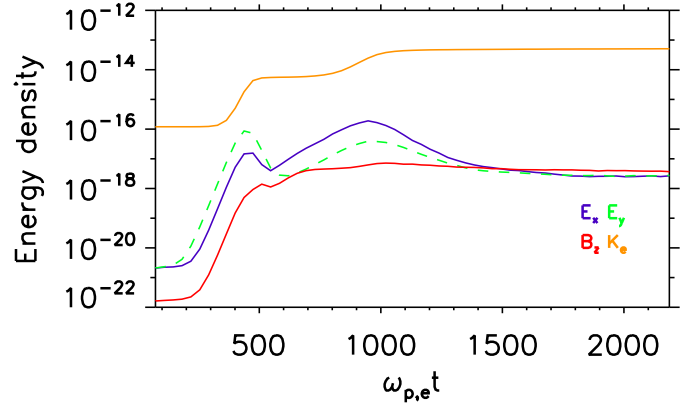}
\caption{Time evolution of the energy densities of electric, magnetic field, and kinetic energy of IGM, respectively, in SI units for run 4.}
\label{fig12}
\end{figure}

\begin{figure}
\includegraphics[width=90mm]{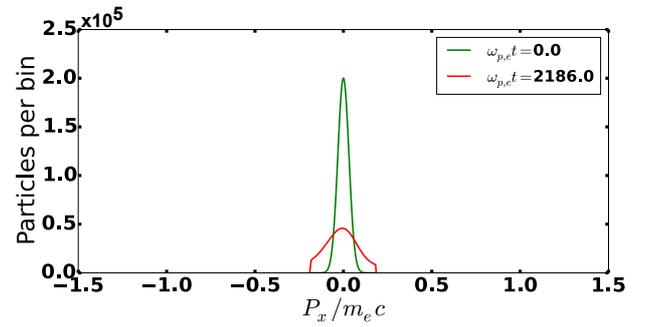}
\caption{IGM momentum distribution in $p_x$ for run 4 at two points in time.}
\label{fig13}
\end{figure}

\begin{figure}
\includegraphics[width=90mm]{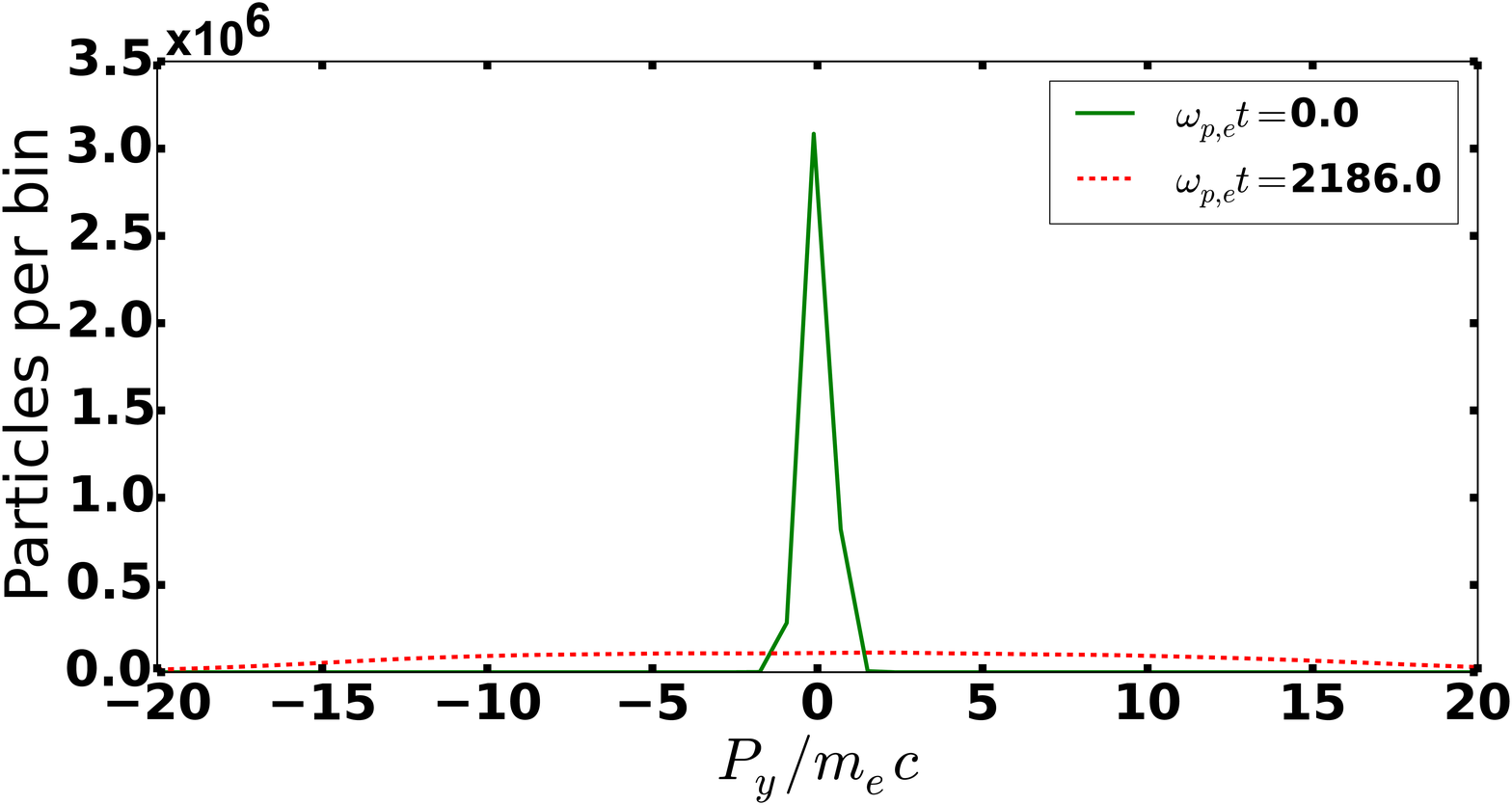}
\caption{Beam momentum distribution in $p_y$ for run 4 at two points in time.}
\label{fig14}
\end{figure}

\begin{figure}
\includegraphics[width=90mm]{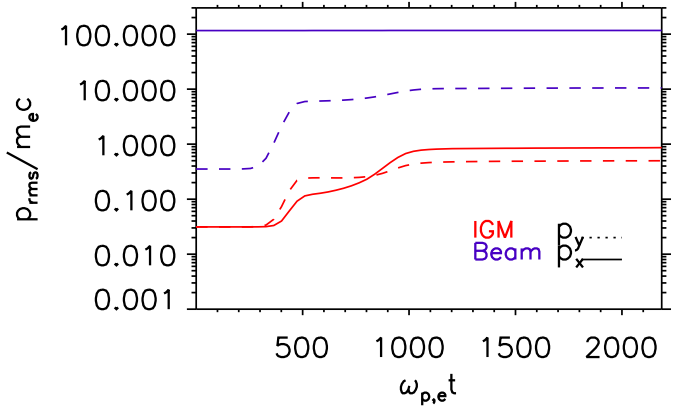}
\caption{Time evolution of the momentum spread of the beam, $p_\mathrm{rms}$, for run 4.}
\label{fig15}
\end{figure}

Finally, run 4 considers a situation in which all three constraints on the beam parameters are violated. Unlike runs 1 and 2, the fastest electrostatic mode develops for wave vectors that are quasi-perpendicular to the beam, as is well seen in Fig. \ref{fig11}. The numerical growth rate perfectly agrees with the analytical estimation for a cold beam, and it is about $2.2\times10^{-2}\omega_{pe}$ which is larger than in the runs 1, 2, and 3. Furthermore, Fig. \ref{fig12} demonstrates that the electric-field energy density assumes a considerably higher value than in three other runs on account of a higher growth rate. At the same time, the Weibel mode is destabilized resulting in a strong growth of magnetic field to a field strength even larger (see the red line in Fig. \ref{fig12}) than that of the electric field. Actually, the orange line in Fig. \ref{fig12} indicates that the dominant energy transfer is that to IGM electrons ($\sim 0.5\%$), while the magnetic field receives only $\sim 10^{-5}\%$.

This affects the momentum distribution of both the beam and IGM, as we present in Figs. \ref{fig13}-\ref{fig14}. In Fig. \ref{fig15} we also see a remarkable increase in the momentum spread of the beam and the IGM. This run 4 is similar to the simulations by \citet{Sironi14} who observed a similar beam-plasma evolution.

\section{Summary}\label{Summary}

We have revisited the issue of plasma instabilities induced by electron-positron beams in the fully ionized intergalactic medium. This problem is related to pair beams produced by TeV radiation of blazars. The main objective of our study is to clarify the feedback of the beam-driven instabilities on the pairs. 

The largest difficulty is the impossibility to simulate realistic blazar-induced beams, even with modern computational resources. Therefore, parameters must be found that permit numerical modeling with similar physical properties. Two important criteria of the realistic pair beams have been noticed before: (i), the beam/IGM energy density ratio is much smaller than unity \citep{Kempf16}, and (ii), the electrostatic mode evolves in the kinetic regime \citep{Miniati13}. However, the simple estimation presented in the introduction shows that the Weibel mode can potentially compete with the kinetic electrostatic instability. To clarify this point, we have used a simple analytical model and demonstrated that the Weibel mode is actually stable for realistic parameters. This adds a third criterion for the pair beams. 

Previous PIC studies of the blazar-induced pair beams \citep{Sironi14,Kempf16} considered only some of these requirements on the beam-plasma system. In contrast, we have performed a simulation (run 1), for which all of them are taken into account. Then, we have compared this case with three other simulations (runs 2-4), for which some criteria were violated. 
The results of run 1 indicate that the pair beam does not experience any significant modification. The electrostatic growth rate turns out to be quite small, and non-linear effects stabilize the beam very efficiently. However, once the electrostatic instability becomes reactive (runs 3-4), as is the case for the studies of \citet{Kempf16} and \citet{Sironi14}, the beam momentum distribution widens drastically in the transverse direction. A significant widening of the beam could in principle account for the observed low flux of cascade gamma rays in the GeV band on account of temporal smearing, but that requires a widening by a factor $\gg 10$. In any case beam widening is only observed if the instability develops in the reactive regime, and that is not relevant for realistic pair beams arising from interactions of AGN gamma rays with extragalactic background light. Also, if the beam/IGM energy density ratio is high, then the beam effectively heats the IGM (run 4), as was seen in the simulations by \citet{Sironi14}. 

To summarize, we have improved modeling of plasma instabilities for blazar-induced pair beams by including three relevant criteria for the beam. Our results suggest that such instabilities play a negligible role and cannot suppress the flux of cascade gamma rays in the GeV band. Thus, other suppression mechanisms of the energy flux from TeV blazars such magnetic-field deflection must be at play. 

\begin{acknowledgements}
The numerical simulations were performed with the EPOCH code that was in part funded by the UK EPSRC grants EP/G054950/1, EP/G056803/1, EP/G055165/1 and EP/ M022463/1. The numerical work was conducted on resources provided by The North-German Supercomputing Alliance (HLRN) under project bbp00003. M.P. acknowledges support through grant PO 1508/1-2 of the Deutsche Forschungsgemeinschaft. The work of J.N. is supported by Narodowe Centrum Nauki through research project DEC-2013/10/E/ST9/00662.
\end{acknowledgements}


\section{Appendix A: derivation of the dispersion equation for $p_{\perp,b}=p_{\parallel,p}=p_{\perp,p}=0$ and $k_\parallel=0$}

In the case $p_{\perp,b}=p_{\parallel,p}=p_{\perp,p}=0$, the beam and plasma distributions, respectively, reads

\begin{multline}
f_b({\bf p})= {n_b\over (p_{\parallel,b}^{+}-p_{\parallel,b}^{-})} \delta(p_z) \delta(p_y) \times \\ \times \l[ \theta\l(p_x-p_{\parallel,b}^{-}\r)-\theta\l(p_x-p_{\parallel,b}^{+}\r) \r] ,
\label{A1}
\end{multline}

\be
f_p({\bf p})= n   \delta(p_x) \delta(p_y)\delta(p_z) ,
\label{A2}
\ee
where $\delta(x)$ is the Dirac delta function. The dielectric tensor is given by \citet{Breizman90} and \citet{RS04}:

\begin{multline}
\epsilon_{i,j}=\delta_{i,j}+ \sum_{a=p,b} {4\pi e^2\over\omega^2}\times \\ \times \int d^3p \l( v_i {\pa f_a({\bf p})\over\pa p_j } - {v_iv_jk_l \over {\bf k}{\bf v}-\omega} {\pa f_a({\bf p}) \over \pa p_l} \r).
\label{A3} 
\end{multline}

Evaluating the dielectric tensor (\ref{A3}) for the distribution functions (\ref{A1})-(\ref{A2}) and for the wave vector ${\bf k}=(0,0,k)$ yields

\be
\epsilon_{zy}=\epsilon_{yz}=\epsilon_{yx}=\epsilon_{xy}=0,
\label{A4}
\ee

\be
\epsilon_{zz}=\epsilon_{yy}=1-{\omega_p^2\over\omega} -{\omega_b^2\over\omega^2}U_1 ,
\label{A5} 
\ee

\be
\epsilon_{xx}=1- {\omega_p^2\over\omega^2}-{\omega_b^2\over\omega^2} 
\l[ \l(kc\over\omega \r)^2 U_1   +  \l( 1- \l(kc\over\omega \r)^2\r)U_2  \r],
\label{A6}
\ee

\be
\epsilon_{xz}=\epsilon_{zx}= - {\omega_b^2\over\omega^2}{kc\over\omega}U_3,
\label{A7}
\ee

where
\be
U_1= {m_ec\over p_{\parallel,b}^{+}-p_{\parallel,b}^{-}}
\ln \l| p_{\parallel,b}^{+}+ [ (p_{\parallel,b}^{+} )^2 +m_e^2c^2]^{1/2} \over p_{\parallel,b}^{-}+ [ (p_{\parallel,b}^{-})^2 +m_e^2c^2]^{1/2} \r|,
\label{A8}
\ee

\be
U_2= {m_ec\over p_{\parallel,b}^{+}-p_{\parallel,b}^{-}}\l( {p_{\parallel,b}^{+}\over [ (p_{\parallel,b}^{+})^2 +m_e^2c^2]^{1/2} } - {p_{\parallel,b}^{-}\over [ (p_{\parallel,b}^{-})^2 +m_e^2c^2]^{1/2} }\r),
\label{A9}
\ee

\be
U_3={m_ec\over 2(p_{\parallel,b}^{+}-p_{\parallel,b}^{-})} \ln\l| (p_{\parallel,b}^{+})^2 +m_e^2c^2 \over (p_{\parallel,b}^{-})^2 +m_e^2c^2 \r|.
\label{A10}
\ee

Here, we have introduced $\omega_b^2=8\pi n_be^2/m_e$. The dispersion equation reads:

\begin{multline}
\det (\Lambda_{i,j})=\det \l(\epsilon_{i,j}+ {k_ik_jc^2\over\omega^2}- \l(kc\over\omega \r)^2\delta_{i,j} \r)= \\
=\Lambda_{yy}(\Lambda_{zz}\Lambda_{xx}-\Lambda_{zx}^2)=0. 
\label{A11}
\end{multline}

Thus, the dispersion equation for electromagnetic fluctuations is

\begin{multline}
\l[ 1-{\omega_p^2\over\omega} -{\omega_b^2\over\omega^2}U_1 \r]
\l\{  1- \l(kc\over\omega \r)^2 - {\omega_p^2\over\omega^2}- \r. \\ \l. -{\omega_b^2\over\omega^2} \l[ \l(kc\over\omega \r)^2 U_1   +  \l( 1- \l(kc\over\omega \r)^2\r)U_2  \r] \r\}- \\ - \l({\omega_b^2\over\omega^2}{kc\over\omega}U_3 \r)^2=0.
\label{A12} 
\end{multline}

\section{Appendix B: derivation of the dispersion equation for $p_{\parallel,b}=0$ and $k_\parallel=0$}

For $p_{\parallel,b}=0$, the distribution function of the beam reads

\begin{multline}
f_b({\bf p})= {n_b\over 4p_{\perp,b}^2 } \l[ \theta\l(p_z+p_{\perp,b}\r)-\theta\l(p_z-p_{\perp,b} \r) \r] \times \\ \times \l[ \theta\l(p_y+p_{\perp,b}\r)-\theta\l(p_y-p_{\perp,b} \r) \r] \delta(p_x-p_0). 
\label{B1}
\end{multline}
We will assume $p_{\perp,b}\ll p_0$. We will still model background protons with the distribution (\ref{A2}), whereas the distribution function of the background electrons is given by Eq. (\ref{an3}). Moreover, we will assume that the background electrons are non-relativistic, $p_{\parallel,p}=m_e v_{\parallel,p}$ and $p_{\perp,p}=m_e v_{\perp,p}$. Now, it is easy to find that again $\epsilon_{zy}=\epsilon_{yz}=\epsilon_{yx}=\epsilon_{xy}=0$, but 

\be
\epsilon_{zz}= 1 - {\omega_{p,p}^2\over\omega^2}- {\omega_{p,e}^2 \over \omega^2- (kv_{\perp,p})^2} - {\omega_b^2/\Gamma\over \omega^2 - (ku)^2},
\label{B2} 
\ee

\begin{multline}
\epsilon_{yy}= 1- {\omega_p^2+\omega_b^2/\Gamma \over\omega^2} - {(kv_{\perp,p})^2 \omega_{p,e}^2\over 3\omega^2( \omega^2 - (kv_{\perp,p})^2 )   } - \\ 
- { (ku)^2\omega_b^2/\Gamma \over 3\omega^2 (\omega^2 - (ku)^2 )} ,
\label{B3} 
\end{multline}

\begin{multline}
\epsilon_{xx}= 1 - {\omega_p^2 + \omega_b^2/\Gamma^3 \over \omega^2} -  {(kv_{\parallel,p})^2 \omega_{p,e}^2\over 3\omega^2( \omega^2 - (kv_{\perp,p})^2 )   } - \\ 
- {(kV_0)^2\omega_b^2/\Gamma \over \omega^2 (\omega^2 - (ku)^2)},
\label{B4} 
\end{multline}

\be
\epsilon_{xz}=\epsilon_{zx}= - {kV_0\omega_b^2/\Gamma \over\omega(\omega^2- (ku)^2 )},
\label{B5}
\ee
where $\omega_{p,e}=(4\pi ne^2/m_e)^{1/2}$, $\omega_{p,p}=(4\pi ne^2/m_p)^{1/2}$ ($m_p$ is the proton mass), $u=V_0p_{\perp,b}/p_0$.

Finally, the Weibel instability is described by the equation

\begin{multline}
\l[ 1 - {\omega_{p,p}^2\over\omega^2}- {\omega_{p,e}^2 \over \omega^2- (kv_{\perp,p})^2} - {\omega_b^2/\Gamma\over \omega^2 - (ku)^2} \r] \times \\ \times 
\l[ 1 - {(kc)^2+\omega_p^2 + \omega_b^2/\Gamma^3 \over \omega^2} -  {(kv_{\parallel,p})^2 \omega_{p,e}^2\over 3\omega^2( \omega^2 - (kv_{\perp,p})^2 )   } - \r. \\ \l. 
- {(kV_0)^2\omega_b^2/\Gamma \over \omega^2 (\omega^2 - (ku)^2)}
\r] - \l[ {kV_0\omega_b^2/\Gamma \over\omega(\omega^2- (ku)^2 )} \r]^2=0.
\label{B6} 
\end{multline}

\section{Appendix C: approximation for $\alpha_{kin}$ at large values of $\mu_R$}
For $\mu_R\gg1$, we can use the series expansion 
\be
c(p^2+m_e^2c^2)^{1/2}- V_0p_x \approx {mc^2\over\Gamma}+ {(p_x-p_0)^2\over 2m_e\Gamma^3} + {{p_z^2+p_y^2}\over 2m_e\Gamma} .
\label{ca1}
\ee
Then Eq. (\ref{k3}) can be approximated as \citep{Watson60,Meierovich76}
\be
f_b({\bf p})= {n_b\over \pi^{3/2} p_{\perp,b}^2 p_{\parallel,b}}e^{-{{p_z^2+p_y^2}\over p_{\perp,b}^2}- {(p_x-p_0)^2\over p_{\parallel,b}^2}}
\label{ca2}
\ee
or
\be
f_b({\bf v})= {n_b\over \pi^{3/2} v_{\perp,b}^2 v_{\parallel,b}}e^{-{{v_z^2+v_y^2}\over v_{\perp,b}^2}- {(v_x-v_0)^2\over v_{\parallel,b}^2}},
\label{ca3}
\ee
where $p_{\perp,b}^2=2 m_ek_BT_{b}$, $p_{\parallel,b}^2=2\Gamma^2 m_ek_BT_{b}$, $v_{\perp,b}^2=2k_BT_{b}/(m_e\Gamma^2)$, $v_{\parallel,b}^2=2k_BT_{b}/(m_e\Gamma^4)$. It is easy to find that
\be
 p_{\perp,b}=(2-\pi/2)^{1/2} {mc\over \mu_R^{1/2}}
\label{ca3.1}
\ee
\be
v_{\parallel, b}=\l( \langle \l(v_x-\langle v_x\rangle \r)^2 \rangle \r)^{1/2}\approx \l( k_B T_{b} \over m_e\Gamma^4\r)^{1/2} 
\label{ca4}
\ee
and 
\be
 \alpha_{kin}= {1\over \Gamma^3 \mu_R^{3/2}}.
\label{ca5}
\ee


\end{document}